\documentclass{IOS-Book-Article}

\usepackage{mathptmx}
\usepackage{url}
\usepackage{graphicx}
\usepackage[table]{xcolor}
\usepackage[table]{xcolor}   
\usepackage{multirow}        
\begin{document}
\begin{frontmatter}              

\title{Can LLMs Create Legally Relevant Summaries and Analyses of Videos?}
\runningtitle{LLM Video Analysis}

\author[A]{\fnms{Lyra} \snm{Hoeben-Kuil},
\thanks{Corresponding Author: Lyra Hoeben-Kuil, Maastricht University, \\ lyra.hoeben-kuil@maastrichtuniversity.nl}
},
\author[A,B]{\fnms{Gijs} \snm{van Dijck}},
\author[C]{\fnms{Jaromir} \snm{Savelka}},
\author[A]{\fnms{Johanna} \snm{Gunawan}},
\author[A]{\fnms{Konrad} \snm{Kollnig}},
\author[A]{\fnms{Marta} \snm{Kołacz}},
\author[A]{\fnms{Mindy} \snm{Duffourc}},
\author[B]{\fnms{Shashank} \snm{M Chakravarthy}},
\author[A]{\fnms{Hannes} \snm{Westermann}}

\runningauthor{Hoeben-Kuil et al.}
\address[A]{Maastricht Law and Tech Lab, Faculty of Law, Maastricht University, The Netherlands}
\address[B]{Brightlands Institute for Smart Society, Maastricht University, The Netherlands}
\address[C]{Computer Science Department, Carnegie Mellon University, Pittsburg, USA}

\begin{abstract}
Understanding the legally relevant factual basis of an event and conveying it through text is a key skill of legal professionals. This skill is important for preparing forms (e.g., insurance claims) or other legal documents (e.g., court claims), but often presents a challenge for laypeople. Current AI approaches aim to bridge this gap, but mostly rely on the user to articulate what has happened in text, which may be challenging for many. Here, we investigate the capability of large language models (LLMs) to understand and summarize events occurring in videos. We ask an LLM to summarize and draft legal letters, based on 120 YouTube videos showing legal issues in various domains. Overall, 71.7\% of the summaries were rated as of high or medium quality, which is a promising result, opening the door to a number of applications in e.g. access to justice.
\end{abstract}

\end{frontmatter}

\thispagestyle{empty}
\pagestyle{empty}

\section*{Introduction}
Many people lack adequate access to justice, leaving them unable to resolve their legal problems \cite{currie2009legal}, with significant costs to both individuals and society \cite{semple2015cost}. A central difficulty lies in the translation of lived experience into (legally) relevant text narratives \cite{westermann2023bridging}. Laypeople often have difficulties identifying which facts are legally relevant and how to put the facts into a legally valid textual representation \cite{branting2020judges,susskind2019online}. Whether in insurance claims, court submissions, or landlord–tenant and consumer disputes, individuals must present facts in a structured, legally meaningful way. The difficulty of people in crafting narratives from their experiences can present an important barrier in obtaining access to justice and in properly utilizing governmental resources \cite{salyzyn2016literacy, branting2020judges, macfarlane2013national}.

Researchers have explored the use of AI to bridge the gap between a layperson understanding of events and a legally relevant explanation \cite{westermann2023bridging,goodson2023intention,steenhuis2024gettingdoorstreamliningintake,branting2023narrative}. However, these approaches still rely on the user interacting with the AI system through text. This means that laypeople who do not have the linguistic skills or vocabulary to describe their issue or identify the relevant details may be left behind. 

In this paper, we explore if multi-modal large language models (LLMs) could analyze and derive legally relevant summaries and legal implications from \textit{videos}. Legal relevancy here refers to the extent to which information, facts, or observations contained in the source (a video) contribute to identifying, interpreting, or evaluating issues that have legal implications. If LLMs have this capacity, it could enable laypeople to simply record a video of their legal issue (such as deficiencies in rented apartments, traffic incidents or product malfunctions) and then use an LLM to fill out a form or draft a legal letter. Thus, the barrier to receive legal help for their situation from AI can be further reduced. 

To investigate this capability of LLMs, we collect a dataset of videos showing a legal issue from YouTube, and set up an evaluation framework to explore the capacity of the models to process the data. We aim to answer the following research questions:

\begin{itemize}
    \item[RQ1] Can LLMs reliably summarize legally relevant facts from video footage?
    \item[RQ2] How do video characteristics (e.g., domain, perspective, voice-over, text, and complexity) influence the quality of LLM outputs?
    \item[RQ3] To what extent can LLMs link these facts to the relevant legal rules (i.e. the legal provisions that apply to the matter at hand)?
\end{itemize}

\noindent The contributions of this paper are thus as follows:
\begin{enumerate}
    \item As far as we are aware, the first study exploring the legal analysis of videos using LLMs.
    \item A dataset containing a collection of video snippet showing legal issues, along with metadata such as domain, complexity and human-written summaries.
    \item A quantitative and qualitative analysis of the capacity of LLMs in summarizing videos and linking them to relevant laws, as well as the factors affecting the quality of the generated text.
    
\end{enumerate}







\section{Related Work}

LLMs have been actively explored in the task of extracting information from video \cite{huang2024vtimellm,shu2023audio}. Alayrac et al. introduced Flamingo---an approach where the model would accept sequences of video frames and perform open-ended tasks such as question answering and captioning \cite{alayrac2022flamingo}. Zhang et al. introduced Video-LLaMA for video comprehension through identification of temporal change in visual scenes \cite{zhang2023video}. 
Maaz et al. introduced Video-ChatGPT which is a form of video-based conversation that merges a video-adapted visual encoder with an LLM trained on a diverse dataset comprising of 100,000 video-instruction pairs \cite{maaz2023video}. 
Lee et al. proposed a novel video summarization framework, called LLM-based video Summarization (LLMVS) which translated video frames into sequences of captions using a Multi-modal LLM \cite{lee2025video}.
Earlier works in video summarization, focusing on courtroom video recordings, utilized unsupervised learning by extracting semantic information through automatic speech transcriptions, automatic audio and video annotations \cite{fersini2009multimedia}. 

In the legal domain, researchers have extensively explored LLMs' abilities in several key areas such as legal reasoning \cite{https://doi.org/10.48550/arxiv.2212.01326,katz2023gpt,katz2024gpt,blair2023can,nguyen2023blackbox}, or document analysis and annotation \cite{savelka2023can,gray2024empirical,savelka2023explaining}. Similarly, access to justice initiatives have evolved significantly from the work such as Protection Order Advisory system \cite{branting2001advisory}, web-based legal decision support systems \cite{zeleznikow2002using} or more recent approaches like the JusticeBot methodology \cite{westermann_justicebot_2023,Westermann_thesis_2023}, combining case-based and rule-based reasoning to help laypeople explore their legal rights. LLMs have also shown promise in dispute mediation \cite{westermann2023llmediator,tan2024robotsmiddleevaluatingllms} and providing legal information to laypeople \cite{tan2023chatgpt}. Recent research has demonstrated their capability to transform legal articles into structured representations for legal decision support tools \cite{janatian2023text} and create automated forms \cite{steenhuis2023weaving}, as well as facilitating the assessment of eligibility for legal aid \cite{steenhuis2024gettingdoorstreamliningintake}. LLMs have also been successfully applied to discover common patterns in fact descriptions of court opinion documents \cite{drapal2023using,gray2024using}. A promising use-case for LLMs in access to justice lies in assessing layperson narratives \cite{goodson2023intention, westermann2023bridging} or layperson provided-pictures \cite{westermann2024analyzingimageslegaldocuments} to help them in understanding their legal situation or filling in forms \cite{westermanndallma}. Here, we expand on this research to study if LLMs can enable laypeople to provide information for such purposes via video clips.

\section{Proposed Framework}
\label{sec:prop_framework}
We suggest a framework that uses an LLM to analyze the legally relevant aspects of that video. First, we ask the LLM to draft a summary of what happens in the video, featuring the legally essential factual occurrences. Then, we ask the model to draft a legal complaint letter, which refers to Dutch legislation and states which norms were violated and how.

\begin{figure}[h]
  \centering
  \includegraphics[width=0.8\textwidth]{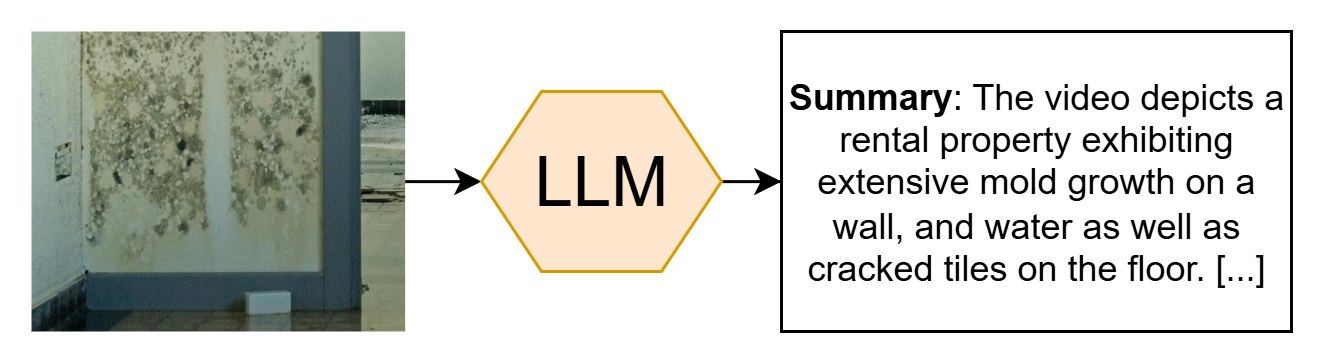}
  \caption{Visualization of the pipeline. Input is a video, output is a summary or a legal complaint letter (not shown).}
  \label{fig:pipeline_summ}
\end{figure}

Such a pipeline could have various practical applications, including for access to justice. Laypeople struggle to know what to do when facing legal issues, or how to articulate what has happened in a legally coherent manner, which can present an important barrier to access to justice, as highlighted by several studies \cite{salyzyn2016literacy,branting2020judges,steenhuis2023beyond,currie2009legal,savage2022experiences}. If LLMs can summarize and analyze video clips, they could help bridge the gap between layperson user and legal support. An app could ask a user to film the problem they face, and then either provide legal information (coupled with a system such as JusticeBot \cite{westermann2023justicebot}) or even take the relevant information and fill out relevant forms or draft letters. Even if not perfect, guiding people to a relevant resource or having a draft that can be reviewed with a lawyer could already be a significant improvement.


Even beyond access to justice, such a system could be very powerful in various AI \& Law applications. Videos represent one of the most abundant and high-fidelity data-sources available to us, and are frequently used as, e.g., evidence. However, until recently, they have mostly been beyond the scope of legal AI systems. We set up experiments to understand the current capacity of LLMs on understanding videos.

\section{Experimental Design}
To evaluate the capability of LLMs to analyze videos from a legal perspective, we collected a dataset of videos (section \ref{sec:dataset}) and built an LLM pipeline to process these videos (section \ref{sec:llm}), in order to generate summaries of the videos (section \ref{sec:e1}) and draft legal letters (section \ref{sec:e2}), which are then evaluated quantitatively and qualitatively. The code to run these experiments is publicly available.\footnote{Available at \url{https://github.com/maastrichtlawtech/jurix2025_LLM_video_analysis}}

\subsection{Dataset}
\label{sec:dataset}

Our dataset consists of 120 videos drawn from four legal domains. All videos were sourced from YouTube, as this platform provides publicly available material that captures how real people record and share footage of disputes and incidents in lived contexts, focused on everyday legal issues \cite{farrow2016everyday}. To ensure comparability across domains, we only included video with at least 2 seconds of footage where a factual legal problem can be seen, and that contained clear visual evidence to allow the factual evaluation.

Videos were selected at random from search results, provided they fit within one of the target domains (see Table \ref{tab:domains}). We also captured metadata for each video (see Table \ref{tab:metadata}).

\begin{table}[h!]
\centering
\renewcommand{\arraystretch}{1.2}
\begin{tabular}{|p{2cm}|p{7.5cm}|c|}
\hline
\textbf{Domain} & \textbf{Description} & \textbf{N Videos} \\
\hline
Housing defects & Focused on everyday disputes and common defects in rental housing. 
Search terms included ``apartment water leak,'',  
``cockroach infestation apartment,'' ``mold in rental apartment,'' 
``landlord refuses repairs,'' ``broken heating.'' & 24 (20\%) \\
\hline
Traffic incidents & Covered a range of vehicle accidents and road user conflicts. 
Search terms included ``car crash compilation,'' ``truck crash dashcam.'' & 42 (35\%) \\
\hline
Property damage & Emphasized natural disasters and destruction. 
Search terms included ``house fire aftermath,'' ``\{storm\textbar flood\textbar earthquake\} damage to my house,'' 
``tenant destroyed property.'' & 34 (28.3\%) \\
\hline
Product malfunctions & Targeted defective consumer goods and devices. 
Search terms included ``batteries randomly catching on fire,'' 
``phone catches on fire,'' ``batteries exploding.'' & 20 (16.7\%) \\
\hline
\textbf{Total} & & \textbf{120 (100\%)} \\
\hline

\end{tabular}
\caption{Domains, search strategies, and number of videos in the dataset.}
\label{tab:domains}
\end{table}

\begin{table}[h!]
\centering
\renewcommand{\arraystretch}{1.3}
\begin{tabular}{|p{8cm}|p{2cm}|c|c|}
\hline
\textbf{Metadata Question} & \textbf{Result} & \textbf{N} & \textbf{\%} \\
\hline
Does the video contain a voice-over? & Yes & 69 & 57.5 \\
                                     & No  & 51 & 42.5 \\
\hline
What is the filming perspective of the video? & First Person & 48 & 40.0 \\
                                              & Third Person & 37 & 30.8 \\
                                              & Dash Cam     & 35 & 29.2 \\
\hline
Does the video contain text in the footage? & Yes & 31 & 25.8 \\
                                            & No  & 89 & 74.2 \\
\hline
\multicolumn{4}{|l|}{\textbf{What is the complexity level of the video?}} \\
\hline
\multicolumn{2}{|p{10cm}|}{\textbf{Level 1}: A single, clearly identifiable event occurs (e.g., a product breaking at a specific moment, water dripping from a ceiling).} & 58 & 48.3 \\
\hline
\multicolumn{2}{|p{10cm}|}{\textbf{Level 2}: Multiple events occur, but they remain individually distinguishable (e.g., a tenant showing several instances of mold damage in sequence).} & 60 & 50.0 \\
\hline
\multicolumn{2}{|p{10cm}|}{\textbf{Level 3}: Many events occur at once, making it difficult to distinguish the legally relevant facts (e.g., a chaotic traffic accident with several vehicles and bystanders).} &  2 &  1.7 \\
\hline

\end{tabular}
\caption{Summary of metadata characteristics across the video dataset.}
\label{tab:metadata}
\end{table}

\subsection{Model}
\label{sec:llm}

To investigate the capabilities of AI to generate factual video summaries, we set up a pipeline that submits short video clips together with task-specific instructions. We used Gemini 2.5 Flash (stable version, last updated June 2025), which represents a state-of-the-art multimodal LLM suitable for processing video inputs. Google enables this model to understand videos by sampling images from the video at 1FPS, and audio at 1Khz. We likewise used functionality that allows the model to directly load videos from Youtube.\footnote{\url{https://ai.google.dev/gemini-api/docs/video-understanding}} 

We interacted with the model via its API, setting the temperature parameter to 0, and seed to 42. The specific prompts used are listed below. As recommended by the provider, we first passed the video to the model and then the textual prompt.

\subsection{E1 - Video Summaries}
\label{sec:e1}
In order to answer RQ1 and RQ2, we used Gemini to write short summaries of a video, to emulate a situation where a layperson has filmed a short clip and wants to use AI to help them fill out information in, e.g., a form. The prompt used was:

\begin{table}[h!]
\renewcommand{\arraystretch}{1.3}
\begin{tabular}{|p{12cm}|}
\hline
\textbf{Prompt:} I will give you a video showing some sort of legal issue. Write a 2-3 sentence summary of the legally relevant elements shown in the video, which will be used as part of a legal form or insurance claim. Make sure to only include facts that are part of the video. Be neutral and specific. \\
\hline
\end{tabular}
\caption{Prompt for Experiment 1 - Creation of video summaries}
\label{tab:prompt1}
\end{table}

Once the summaries were generated, we assessed their quality through an annotation process. We drafted annotation guidelines and asked annotators (co-authors of this paper) to annotate each video with the  criteria explained in Table \ref{tab:results_overall}.
In total, seven annotators worked on the annotation, and each video was annotated by two. Annotators were instructed to ignore external knowledge, assumptions, or legally relevant inferences that were not explicitly visible in the video. In case of disagreements, a single annotator reconciled the annotations in taking into account the notes written by the initial annotators.

\subsection{E2 - Legal letter drafts}
\label{sec:e2}
To answer RQ3, we also wanted to assess whether an LLM can flag legally relevant issues and link them to relevant legislation. This mimics a situation where a layperson asks an LLM to draft a full legal complaint letter based on a video. We selected 5 videos from each category and asked the model to draft a letter under dutch law. The prompt can be seen in Table \ref{tab:prompt2}. To ensure consistency and legal relevance, we specified who the addressee of the letter should be for each domain, see table \ref{tab:addressee}.
The resulting complaint letters were then qualitatively analyzed by legal experts to determine whether the model correctly identified and articulated core aspects of legal reasoning, such as identification of violated norms, damages and remedies claimed.


\begin{table}[h!]
\renewcommand{\arraystretch}{1.3}
\begin{tabular}{|p{12cm}|}
\hline
\textbf{Prompt:} Assume that the video happens in the Netherlands. Write a legal complaint letter from the perspective of the person filming the video, referring to relevant legal rules in the Netherlands. Write the letter in English. Address the letter to the \{addressee\}. \\
\hline
\end{tabular}
\caption{Prompt for Experiment 2 - Drafting of legal letters}
\label{tab:prompt2}
\end{table}

\begin{table}[h!]
\centering
\renewcommand{\arraystretch}{1.2}
\begin{tabular}{|p{4cm}|p{8cm}|}
\hline
\textbf{Domain} & \textbf{Addressee} \\
\hline
Housing defects & Landlord \\
\hline
Property damage & Person who caused the damage \\
\hline
Traffic & Car that caused the crash \\
\hline
Product malfunction & Product producer \\
\hline
\end{tabular}
\caption{Mapping of domains to the relevant addressee.}
\label{tab:addressee}
\end{table}

\section{Results}
\subsection{E1 - Video Summaries}
\begin{figure}[h!]
    \centering
    \includegraphics[width=\textwidth]{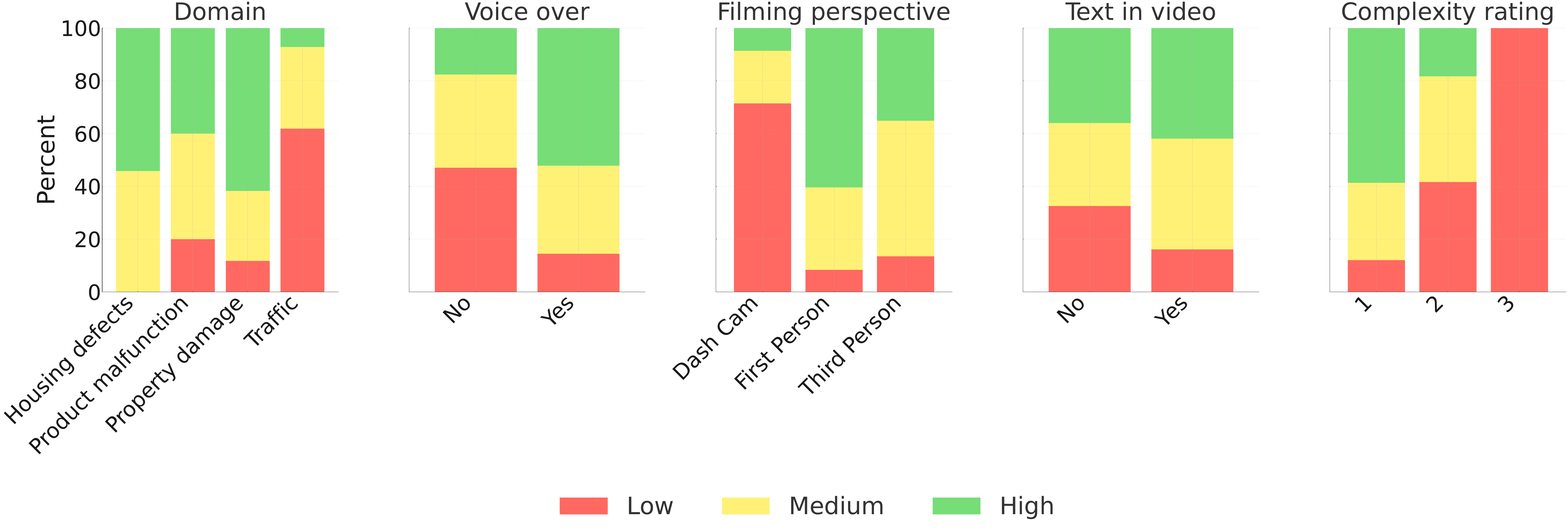}
    \caption{Overall quality of AI summaries across metadata factors (domain, voice-over, perspective, text, complexity).}
    \label{fig:overall_quality_by_metadata}
\end{figure}

\begin{figure}[h!]
    \centering
    \includegraphics[width=\textwidth]{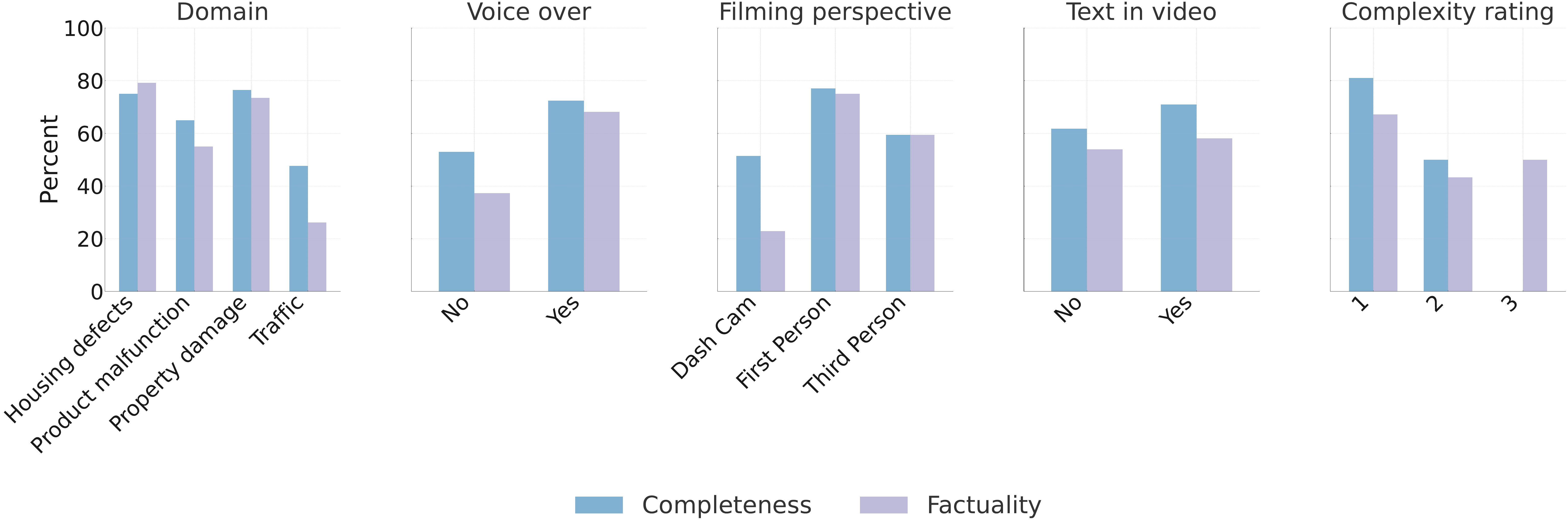}
    \caption{Completeness and factuality of AI summaries across metadata factors (domain, voice-over, perspective, text, complexity).}
    \label{fig:completeness_factuality}
\end{figure}

To answer RQ1 and RQ2, we generated summaries for 120 video clips and evaluated them based on overall quality, completeness and factuality. Table \ref{tab:results_overall} shows the overall results of the rated video summaries. Overall, 37.5\% of the summaries were rated as being of ``high'' quality, while 34.2\% were ranked as ``medium'' and 28.3\% were rated as being of low quality. 64.2\% were found to be complete, while 55\% were found to be factual. 
Figure \ref{fig:overall_quality_by_metadata} shows how the quality of the generated summaries varies by the various metadata (see section \ref{sec:dataset}). Figure \ref{fig:completeness_factuality} shows how the completeness and factuality metrics vary based on the metadata.

\begin{table}[h!]
\centering
\renewcommand{\arraystretch}{1.3}
\begin{tabular}{|p{9cm}|c|c|}
\hline
\textbf{Question / Criteria} & \textbf{N} & \textbf{\%} \\
\hline
\multicolumn{3}{|l|}{\textbf{Overall Quality of AI Summary}} \\
\hline
\multicolumn{1}{|p{9cm}|}{\cellcolor{green!25}\textbf{High}: All facts captured, no hallucinations.} & 45 & 37.5 \\
\hline
\multicolumn{1}{|p{9cm}|}{\cellcolor{yellow!25}\textbf{Medium}: Some facts missing or small additions, but without essential impact on the meaning of the footage.} & 41 & 34.2 \\
\hline
\multicolumn{1}{|p{9cm}|}{\cellcolor{red!25}\textbf{Low}: Key facts missing and/or hallucinations that misrepresent the event and drastically change its meaning.} & 34 & 28.3 \\
\hline
\multicolumn{3}{|l|}{\textbf{Completeness: Were all legally relevant occurrences visible in the video captured in the summary?}} \\
\hline
\multicolumn{1}{|p{9cm}|}{\cellcolor{green!25}\textbf{Yes}: The summary captured all legally relevant occurrences.} & 77 & 64.2 \\
\hline
\multicolumn{1}{|p{9cm}|}{\cellcolor{red!25}\textbf{No}: Some legally relevant occurrences were missing.} & 43 & 35.8 \\
\hline
\multicolumn{3}{|l|}{\textbf{Factuality: Were all of the facts in the description accurate to the video (hallucination-free)?}} \\
\hline
\multicolumn{1}{|p{9cm}|}{\cellcolor{green!25}\textbf{Yes}: The summary was factually accurate and hallucination-free.} & 66 & 55.0 \\
\hline
\multicolumn{1}{|p{9cm}|}{\cellcolor{red!25}\textbf{No}: The summary contained hallucinated or factually incorrect details.} & 54 & 45.0 \\
\hline
\end{tabular}
\caption{Overall quality, completeness, and factuality of AI-generated summaries.}
\label{tab:results_overall}
\end{table}

\newpage\subsection{E2 - Legal Letter Drafts}
To answer RQ3, we used an LLM to draft legal letters based on 5 videos for each domain, (20 in total) and qualitatively evaluated the results. In general, the letters demonstrated structural clarity, references to relevant provisions, and a degree of legal reasoning that makes them functional for preliminary use. However, certain limitations emerged, particularly with respect to contextual accuracy and consistency in the application of legal standards. A more detailed analysis of these results will be presented in Section \ref{Section 5.3}.


\section{Discussion}

Let us examine what the results imply about the research questions.

\subsection{RQ1 - Can LLMs reliably summarize legally relevant facts from video footage?}

E1 indicates that the video summaries are promising, with 71.7\% of the summaries being rated as of high or medium quality. Overall, 64.2\% of the generated summaries captured all relevant occurrences, while 55\% of the videos were found to be free of hallucinations. 

Considering the difficulty of the task of observing a video with sound and summarizing it in a coherent manner that captures the relevant legal details, these results are promising. However, they also highlight that the tool could not reliably generate accurate summaries. If deployed today, one would need a human in the loop to correct them. That said, being presented with a draft summary could reduce the time needed to draft a legal submission for lawyers, and present a useful first draft even for laypeople, who could correct factual mistakes but benefit from the tone and wording of the summaries.

We would also like to highlight the difficulty of annotating the videos. In the end, many of the questions saw significant disagreement between the annotators. Quality and completeness saw especially high disagreements. Partially, this may be resolved with more precise instructions of what should be covered in each summary or a different prompting style, which we will explore in future work. However, the question of which fact in a video is ``legally relevant'' is also quite ambiguous, and may lead to different answers between different legal experts as well. Once more, this research points to the fact that LLMs are at a level where evaluation against a gold-standard dataset is starting to become very difficult (see also \cite{ma2023conceptual,kapoor2024promises,tan2024robotsmiddleevaluatingllms, janatian2023text}).

\subsection{RQ2 - How do video characteristics influence the quality of LLM outputs?}


\textbf{Domain}: The domain had a large impact on the quality of the summaries (Figure \ref{fig:overall_quality_by_metadata}):

\textit{Housing defects:} Housing defects had the highest score out of all of the domains. Regarding completeness, the generated summaries were not specific enough in e.g. the size of mold or that water was wastewater. The hallucinations centered on being too confident in certain facts (i.e. bugs being cockroaches), spotting things not present in the video (e.g. debris or aerial spores) and over-relying on the narration.

\textit{Product malfunction}: Overall quite reasonable scores. However, the model was bad at understanding sequences of events, especially when in rapid succession (see discussion regarding frame rate below). Here, the summary would make up things happening (such as objects being thrown or people being propelled by an explosion). 

\textit{Property damage}: Second-highest scoring category. The model sometimes over-relied on the voice-over, and sometimes ignored it. It likewise had issues in assessing the height of water level and would often skip over the general messiness of a room. 

\textit{Traffic}: By far the lowest scoring category. The model seems unable to capture sequences of events when the car was moving, and instead fully hallucinates what happens. This could stem from the fact that the model only receives a single frame per second. Future work will need to explore increasing the framerate. Even so, the model would often confuse directions, such as left and right, or passenger-side versus driver-side. Emergency vehicles and signs were also hallucinated in some instances.



\textbf{Voice-over}: Videos with a voice-over had higher quality ratings, completeness and factuality scores. The model was able to understand spoken voices in Dutch as well as English, and weaved the information into the description. In some instances, this caused hallucinations, since the model would believe what the voice said even if it was not backed up in the video, which may not be desirable in all circumstances.

\textbf{Perspective}: Dash cam, which was exclusively used in traffic videos, performed poorly. First person performed a bit better than third-person videos, perhaps because the person in front of the camera adds another element for the LLM to consider.

\textbf{Text}: Just like voice-over, text in videos did improve performance. The summaries often included date and times from overlays, and relied on the text.

\textbf{Complexity}: Videos of higher complexity were  harder for the LLM to accurately summarize. Note that there are only 2 samples of complexity level 3.

\subsection{RQ3 - To what extent can LLMs link these facts to the relevant legal rules?} 
\label{Section 5.3}
In E2, we asked the LLM to go further than merely summarizing the videos, by also drafting a legal complaint letter under Dutch law, which we then qualitatively evaluated. The legal rules differ by domain and letter addressee (see section \ref{sec:e2}).

\subsubsection{Housing Defects}
In the domain of housing defects, AI-generated letters often identified relevant statutory provisions and proposed potential remedies, providing a structured draft suitable for preliminary use. However, significant limitations were evident. The drafts consistently omitted reference to the Rental Defects Guide, a source of soft law that, although not formally binding, is systematically relied upon by courts and rental tribunals in such disputes. Furthermore, while the prescribed response period was often set at 14 days, in line with Dutch case law, some drafts specify significantly shorter deadlines, which would not be upheld in court. In addition, there was variability in the framing of remedies; for instance, the letters often did not specify amounts, which, given the nature of the video evidence, is understandable. Overall, while these drafts may serve as a starting point, they require substantial revision and careful expert oversight before being used in practice.

\subsubsection{Product Malfunction}
In cases involving product malfunctions, AI-generated letters generally relied on strict liability provisions, such as product liability statutes (e.g., Articles 6:185–6:186 DCC), and highlighted core principles such as consumer safety expectations. While the drafts suggested remedies and response deadlines that align with legal practice, a notable limitation was the consistent omission of complementary legal claims, particularly contractual non-conformity under Article 7:17 DCC or the general provision of tort (art. 6:162 DCC). This omission reduces the legal strategy of the letters, as both the producer and the seller could otherwise be held liable. In addition, the damage was not described in detail (no amounts are specified) resulting in letters that request acknowledgment of a compensation for all losses, regardless of scale.

\subsubsection{Property damage}
In property damage cases, the quality of AI-generated letters depended on the intended purpose. When tasked with describing the issue and drafting correspondence for the landlord, the letters demonstrated some functional utility: they referred to relevant legislative provisions and identified possible follow-up actions. However, the drafts did not specify the items that were damaged nor the value of the losses incurred, limiting the precision and practical usability of the letters. Despite these limitations, the outputs may still serve as a useful starting point for discussion and further refinement by legal professionals.

\subsubsection{Traffic}
In traffic-related cases, AI-generated letters again showed structural clarity and the ability to correctly identify and reference statutory provisions. However, factual inaccuracies were common, and because legal analysis relies on the stated facts, these errors directly undermined the correctness of the legal reasoning. This illustrates how deficiencies in factual accuracy ``bleed through'' to the legal assessment, emphasizing the necessity of thorough human review. Consequently, the drafts are largely unusable. Additionally, the letters tended to focus on alleged traffic law violations, which may appear intuitively relevant but are legally inappropriate as a basis for civil claims between private individuals; whether a violation occurred is determined by criminal court. 

\subsubsection{Conclusion}
Overall, the draft letters demonstrate structural clarity, the ability to reference relevant statutory provisions, and preliminary legal reasoning, making them potentially useful as starting points for drafting. However, usability is limited by recurring shortcomings, including factual inaccuracies, omissions of complementary legal claims, inconsistent or incomplete specification of damages, the omission of relevant soft law, and variability in proposed response periods. In particular, factual inaccuracies “bleed through” to the legal assessment, undermining the validity of the reasoning in mostly traffic disputes. This indicates that, while AI can assist in the  drafting process and may be useful as a starting point for discussion, careful human review is still essential. 

\subsection{Toward a decision support app analyzing videos}
Finally, let us explore the question of whether and how the current state of video analysis using LLMs could be used in a legal decision support tool, as described in section \ref{sec:prop_framework}.

Despite the unreliability of the generated summaries, we think they could serve as a useful step to drafting a description, especially in domains such as housing defects. The user should be informed to take a stable video that clearly shows the issue, and potentially describe what they see as well. Ideally each clip should show a single issue. 

Even so, the text should be viewed and corrected by a legal professional, before being entered into a final form. The tool may likewise be used as an aid for identifying potential eligibility for compensation or legal recourse. In this capacity, it could help users recognize when they might have a valid claim and guide them toward appropriate next steps, such as contacting legal aid services or consumer protection organizations.

Regarding the drafting of legal letters, the models were able to draft legally sound content drawing from Dutch law, which is a quite small jurisdiction. However, the content was not perfect and also suffered from the inaccuracy in summarizing the videos. Integrating the video-recognition system into an existing legal information pipeline, such as the approach shown in \cite{westermanndallma}, may address some of these shortcomings.


\section{Conclusion \& Future Work}
We investigated the ability of LLMs to summarize videos from a legal perspective, and draft legal letters. The results are promising, but have room for improvement in terms of completeness and factuality, especially for complex videos with significant movement. We explored ways to mitigate the issues and opportunities for access to justice.

Future work will focus on enhancing the prompting (e.g. through chain-of-thought, reasoning models and RAG) and evaluation of the pipeline, experimenting with different video frame-rates, integration in neuro-symbolic systems and user interfaces,  testing with users, and investigating more complex videos covering e.g. financial crimes or events from multiple perspectives.

\bibliographystyle{plain}
\bibliography{biblio}

\end{document}